\documentclass[10pt]{article}
\usepackage{amsmath,bm}
\usepackage{graphicx,pdflscape}
\usepackage[toc,page]{appendix}

\usepackage{color,subfigure}
\usepackage{soul}
\graphicspath{{images/}}
\usepackage{mathtools}
\usepackage{amsmath}
\usepackage{hyperref}
\hypersetup{
    colorlinks,
    citecolor=red,
    filecolor=cyan,
    linkcolor=blue,
   urlcolor=magenta
 }

\graphicspath{{images/}{../}}
\usepackage{epstopdf}

\usepackage{tabularx}

\usepackage{capt-of}

\usepackage{listings}

\usepackage{soul}

\definecolor{dkgreen}{rgb}{0,0.6,0}
\definecolor{gray}{rgb}{0.5,0.5,0.5}
\definecolor{mauve}{rgb}{0.58,0,0.82}

\usepackage{float}
\usepackage{hyperref} 

\usepackage{xparse}
\ExplSyntaxOn
\NewDocumentCommand{\mref}{m}{\quinn_mref:n {#1}}
\seq_new:N \l_quinn_mref_seq
\cs_new:Npn \quinn_mref:n #1
 {
  \seq_set_split:Nnn \l_quinn_mref_seq { , } { #1 }
  \seq_pop_right:NN \l_quinn_mref_seq \l_tmpa_tl
  ( 
  \seq_map_inline:Nn \l_quinn_mref_seq
    { \ref{##1},\nobreakspace } 
  \exp_args:NV \ref \l_tmpa_tl 
  ) 
 }
\ExplSyntaxOff

\newcommand{\sri}[1]{{\color{black} #1}} 
\newcommand{\disp}[1]{Eq.~\mref{#1}}

\newcommand{\figdisp}[1]{Fig.~\mref{#1}}
\newcommand{\tabdisp}[1]{Table~\ref{#1}}

\newcommand{\lessim} {\ {\raise-.5ex\hbox{$\buildrel<\over\sim$}}\ }

\newcommand{\gssim}{\ {\raise-.5ex\hbox{$\buildrel>\over\sim$}}\ }

\newcommand{\beq}{\begin{eqnarray}}
\newcommand{\eeq}{\end{eqnarray}}
\newcommand{\barray}{\begin{eqnarray}}
\newcommand{\earray}{\end{eqnarray}}
\newcommand{\half}{\frac{1}{2}}

\newcommand{\tJ}{\ $t$-$J$ \ }
\newcommand{\nn}{\nonumber}



\usepackage{xcolor}
\colorlet{redd}{red}

\title{Aspects of the normal state resistivity of  cuprate superconductors $Bi2201$, $Tl2201$ and $Hg1201$} 
\author{ Samantha Shears\footnote{sshears@.ucsc.edu},  Michael Arciniaga\footnote{michael.arciniaga@gmail.com} and B Sriram Shastry\footnote{sriram@physics.ucsc.edu}  \\
\small \em $^{1}$Physics Department, University of California, Santa Cruz, CA, 95064 
}

\date{July 2, 2025}

\begin{document}

\maketitle

\section*{Abstract}
Planar normal state resistivity data from three families of hole doped single layer cuprate superconductors   $Bi2201$ (Bi$_2$Sr$_2$CuO$_{6+x}$), $Tl2201$ (Tl$_2$Ba$_2$CuO$_{6+x}$) and $Hg1201$ (HgBa$_2$CuO$_{4+x}$) are calculated using the extremely correlated Fermi liquid theory (ECFL). \sri{This theory was  recently employed by us  for  computing the resistivity of  three families  of single layer cuprate superconductors LSCO, BSLCO and NCCO, followed by a detailed comparison. Adding the three systems studied here,   accounts for  {\em all} the remaining  single layer compounds}, where data is available for a  range of densities and temperatures, \sri{ thereby providing a comprehensive study of one class of important cuprate superconductors. } The added  study of the material  $Bi2201$ is  of particular interest, since it is the  system where the almost linear in temperature resistivity was \sri{first} reported in 1990.     \sri{Only recently, in 2022, has a systematic doping analysis become available}. The $Tl2201$ system has  two distinct sets of  band parameters that fit the same Fermi surface, providing new challenges and  insights into the ECFL theory.


\section{Introduction}
Strongly correlated systems such as high $T_c$ systems provide a formidable  challenge to our current understanding of the physics of interacting Fermi systems. The standard framework is largely built  using the  density functional theory of Kohn {\em et. al}, supplemented by methods incorporating weak or moderate strength  interactions. New techniques to calculate the physics of strongly correlated systems, where interactions are much bigger than the band energy are
few, and their reliability is not fully established. A major question that remains to be settled  is whether such strongly correlated systems  are Fermi liquids, or some variety of non-Fermi liquids.  Monitoring and interpreting  the behavior of the resistivity in the normal state  can- in principle- identify the nature of the underlying normal state and distinguish between Fermi-liquids and non Fermi-liquids. Experimental data on many systems shows a complex set of T dependences  in different regimes, varying with the density, and understanding them from the theoretical viewpoint is our main task. 

The extremely correlated Fermi liquid theory (ECFL) \cite{ECFL} was developed starting in 2011, \sri{ to extend the theory of resistivity due to  inelastic electron-electron scattering. This important mechanism was first suggested by Landau and Pomerantschuk\cite{Landau-Pomeranchuk} in  1937. They noted that electron-electron scattering would  lead to a $T^2$ contribution to  the resistivity of simple metals, and on adding the phonon scattering contribution, giving rise to a $\rho\sim \alpha T^2+ \beta T^5$ behavior at low temperatures. The key point made in \cite{Landau-Pomeranchuk} is that in a simple metal with  electrons moving in a Bloch band,  the conservation of the total  momentum of a pair of electrons in the scattering process does not imply the conservation of their total {\em velocity}.  In the presence of the lattice the conserved object is the ``crystal momentum''- i.e. momentum modulo a reciprocal lattice vector, and the velocity refers to the group velocity of the electron waves. This feature allows a certain fraction of the scattering processes-
 the  {\em umklapp} processes- with a non-zero reciprocal lattice vector to  balance the momentum,  while allowing a non-zero transfer of velocity, leading to a non-vanishing resistivity. This is true even in the case of a single band of electrons, while adding other bands makes this easier. For simple metals the magnitude of the umklapp contribution was quantitatively estimated in 1972 \cite{Lawrence-Wilkins}. In metals  with stronger interactions - such as transition metals and heavy-Fermi systems \cite{MJRice,Miyake-Varma}, the umklapp fraction is estimated to be close to unity.

While strongly correlated metals are experimentally very important and wide spread in occurrence, their theoretical treatment using controlled methods is a highly challenging problem. The usual method of using the interaction  strength as a perturbative parameter is clearly undermined if it is very large, even bigger than the band width, and resorting to summation of diagrams etc becomes very dubious. This has created a large theoretical gap in the space of techniques, wherein the ECFL theory has been launched in 2011 \cite{ECFL}.  The ECFL theory yields for strongly correlated systems a resistivity as a function of the basic material parameters including the band parameters, density, the interaction constants, etc (see \disp{r-formula}). At low $T$ and for densities close to half-filling, the Landau-Pomerantschuk resistivity $\rho\sim T^2$ is replaced by a complex behavior,   with $\rho\sim T^2$ at very low T crossing over to a linear i.e. $\rho \sim T$  behavior. The crossover temperature  is remarkably low for most parameters, as a consequence of strong correlations and the proximity to a Mott-Hubbard insulating state.

}

\sri{ The ECFL theory  is currently formulated for systems that can be described  by a single correlated band. The interactions can be modeled  by the large U-Hubbard model or the closely related \tJ model.  }
The ECFL theory is able to provide quantitative results for the resistivity, using the following ingredients:
\begin{itemize}
\item[(i)] A single copper oxide with dimensionless band parameters $t'/t,t''/t\ldots$ retrieved from the shape of the Fermi surface determined by angle resolved photo emission (ARPES). Here the interlayer hopping is assumed to be negligible.
\item[(ii)] The particle density $n$ (the number of electrons per copper), usually obtainable from the Luttinger-Ward area of the Fermi surface found from ARPES.
\item[(iii)] The interlayer lattice constant $c_0$ obtainable from crystallography- it is usually half the c axis lattice constant $c_L$ in the almost tetragonal unit cell. 
 \end{itemize}
These items determine all  parameters in the \tJ model \disp{tJmodel}, with the exception of  $J$  and $t$ itself. Our earlier  results suggest that $J$ is not a sensitive parameter\cite{J-related} and we take $J/t\sim0.17$ in most of our work. The value of 
$t$ is the single adjustable parameter that is fixed for each family of materials studied, by choosing a reasonable overall fit to the resistivity over many densities. Having access to data sets containing   several densities is advantageous, with an overall fit one  can expect to reduce the implicit bias in the fits  if only   a single density is  considered.  It should be noted that the results of the ECFL (see \figdisp{Bi2201-All,Tl2201-All}) can  be broadly characterized as leading to a resistivity that is quadratic in temperature below a surprisingly low scale (given the large t$\sim$ 1 eV), which crosses over to an almost linear behavior over a wide temperature scale,   often with another  crossover- and finally with sight curvature reappearing at fairly high T ($\sim$ 600 K). The quasiparticle weight turns out to be much reduced from unity, and the crossover T scales are sensitively dependent on the density and   band  parameters $t'/t,t''/t\ldots$. The detailed equations of the ECFL theory given in \cite{ECFL,Aspects-I}, and summarized below, produce this complex variety of behaviour starting from the microscopic parameters defining the model \disp{tJmodel}.

In a recent paper \cite{Aspects-I} we applied the  ECFL to four major families of cuprate superconductors- LSCO\cite{Ando-1}, BSLCO\cite{Ando-1}, NCCO\cite{NCCO} and LCCO\cite{LCCO}-  where all the above ingredients are present. These systems  are 
characterized by a single sheeted Fermi surface and with single layer (i.e. well separated) copper oxide planes, that allow or a quasi 2-dimensional theory to be applied. It is shown in that paper that theory shows quantitative agreements  with experiments  over several densities. For LSCO we studied samples at 11 densities, and for BSLCO we studied samples at 7 densities. For the electron doped materials NCCO we studied the 2 available metallic samples and for LCCO we studied samples at 4 densities. The temperature range of most of the systems was from $T_c$ up to 300  and  400 K in the case of LSCO. In most cases \cite{Aspects-I} reports a  close agreement between theory and experiment. 

\sri{The present study takes the goals of \cite{Aspects-I} forward, by including three other systems, and thus  providing a comparison between the ECFL theory and experiment for  {\em all } single layered cuprates known so far. For this purpose we study  the following  compounds here.}
The  single layer system Bi2201 was omitted from our study in \cite{Aspects-I} since results were available for only a single density at that time\cite{Fiory,Martin}, and is included in this work since further data has been published meanwhile\cite{Hussey}. This system was experimentally studied in a few  influential papers  \cite{Fiory,Martin} in 1989-90. In these papers  T-linearity of resistivity was  reported over a remarkably large range of T, between 8 K and  $\sim$800 K. This  result   was expected  to be a harbinger of universal T-linearity of resistivity in the cuprates, therefore possibly implying the general demise of any kind of Fermi liquid theory in these systems. However the reported results were confined to a single composition, and hence some of the ingredients mentioned above were missing. 
 The situation remained unchanged for almost three decades until very recently. This system has been studied recently  in \cite{Hussey}, who have reported data on a few different densities overlapping with that in \cite{Fiory,Martin}, albeit over a  smaller temperature range $T\lessim 300$K.  
 
 New results on another interesting  single layer system Tl2201 at a set of  densities have also been reported recently in \cite{Tl2201,Cooper1,Cooper2}. This system is of additional interest since it allows convenient access to the highly overdoped regime.
 The present work extends the earlier work \cite{Aspects-I} to include parameters relevant to the available samples of Bi2201 and Tl2201.  
 
  We mention that Tl2201 leads to an interesting and unexpected theoretical situation, we found that  the reported Fermi surface can be fit with a significantly different set of band parameters from the ones reported in  \cite{Cooper1,Cooper2}, and we are able to non-trivially test a  theoretical hypothesis that it is the shape of the Fermi surface- rather than the values of the band parameters- that determine the computed resistivity.  For context we note that in the \tJ model, the hopping parameters multiply the (Gutzwiller) correlated Fermi operators which can be viewed as consisting of 4 Fermions,   and hence this hypothesis seems to require testing.
 
\sri{ Finally we  present the results of  study of the system Hg1201 \cite{Yamamoto,Yamamoto2,Greven,Vishik,TDas,Pelc}. This system together with the two compounds listed above and the four compounds  in \cite{Aspects-I} completes the known list of single layer cuprates. 
}

\subsection{The \tJ model and the ECFL methodology}

The $t$-$J$ model \cite{tJmodel} is very important for understanding strongly correlated systems. This model is related to the Hubbard model in the $U\to \infty$ limit, precluding double occupancy.
The model is written in the usual form
\beq
H=P_G H_{tb} P_G+ J \sum_{<i,j>}(\vec{S}_i.\vec{S}_j- \frac{1}{4} n_i n_j) \label{tJmodel}
\eeq
where the first term is the  Gutzwiller projected  band energy,  i.e. $P_G$ is the Gutzwiller projector, and  the exchange term is restricted to nearest neighbours. The tight binding term is written as a sum over a range of neighbours, where $\vec{r}_i \to i$ are the locations of the lattice sites assumed to be on a square lattice with lattice constant $a_0$ and with      
\beq
H_{tb}= -\sum_{ij} t_{ij} C^\dagger_{i \sigma} C_{j \sigma}= \sum_{k \sigma} \varepsilon_k C^\dagger_{k \sigma} C_{k \sigma}
\eeq
with 
\beq -t_{ij}=- t \delta_{|i-j|=a_0}- t' \delta_{|i-j|=\sqrt{2} a_0}-t'' \delta_{|i-j|=2 a_0} \eeq
and the Fourier transform of $-t_{ij}$ is the band dispersion $\varepsilon(\vec{k})$ given by 
\begin{equation}
    \varepsilon(\vec{k})=-2t (\cos(k_x a_0)+\cos(k_y a_0)) -4 t' \cos(k_x a_0)\cos(k_y a_0) - 2 t''(\cos(2k_x a_0)+\cos(2 k_y a_0)).
    \label{band}
\end{equation}

  Details of the ECFL formalism  has been discussed extensively in prior papers \cite{ECFL}, and also the resistivity related paper \cite{Aspects-I}. Here we will provide the barest overview to familiarize the reader with notations.

In ECFL a one electron Green's function can be broken into the product of an auxiliary Green's function $\mathbf{g}$ and the caparison  function $\Tilde{\mu}$:
\begin{equation}
    G(\vec{k},i\omega_n)=\mathbf{g}(\vec{k},i\omega_n)\times \Tilde{\mu}(\vec{k},i\omega_n)
\end{equation}
with $\omega_n= \frac{2 \pi}{\beta}(n+\half)$ is the fermionic  Matsubara frequency, and $\mathbf{g}(\vec{k},i\omega_n)$ is a canonical fermion propagator. $\Tilde{\mu}$ and $\mathbf{g}$ are found from  two self energies $\Psi(\vec{k},i\omega_n)$ and $\chi(\vec{k},i\omega_n)$
\beq
    \Tilde{\mu}(\vec{k},i\omega_n)&=&1-\lambda\frac{n}{2}+\lambda\Psi(\vec{k},i\omega_n), \nn \\
    \mathbf{g}(\vec{k},i\omega_n)^{-1}&=&{i\omega_n+ \mu' -\Tilde{\mu}(\vec{k},i \omega_n)(\varepsilon(\vec{k})-u_0/2)-\lambda\chi(\vec{k},i\omega_n)},
\eeq
where $\lambda$ is an interpolation parameter  set equal to 1 at the end,  $\mu'=\mu-\half u_0+  \lambda  n J$, and $u_0$ is a Lagrange multiplier, which along with the thermodynamic chemical potential $\mu$  is  fixed  from  two particle number  sum-rules
\beq
    n_G&=&2\Sigma_k G(k)e^{i\omega_n 0^+}=n, \nn \\
    n_g&=&2\Sigma_k \mathbf{g}(k)e^{i\omega_n 0^+}=n.
\eeq
In the ECFL theory the two self energies satisfy coupled integral equations that are available as an expansion in powers of $\lambda$, this is truncated   to second order for this problem as in \cite{Aspects-I}. We note that $\lambda=0$ gives the non-interacting theory, whereas the exact Gutzwiller projected theory requires a summation of the $\lambda$ expansion to all orders. By truncating the expansion to second order we are making an approximation to the exact theory, which captures some of the significant effects of strong correlations, as argued in \cite{ECFL,Aspects-I}.
Solving these equation gives the spectral function $A(\vec{k},\omega)$ found by 
analytically continuing to real frequencies from the Matsubara frequencies $i\omega_n\to\omega+i0^+$ by using $A(\vec{k},\omega)=-\frac{1}{\pi}\text{Im} G(\vec{k},\omega)$.

\subsection{Formulation for resistivity}
Within the ECFL theory we express the resistivity as
\beq
\rho=R_{vK}\times c_0\times \bar{\rho}\left(\frac{t'}{t},\frac{t''}{t},\frac{k_B T}{t},\frac{J}{t},n \right) \label{r-formula}
\eeq
where $R_{vK}=\frac{h}{e^2}=25813 \Omega$ is the von Klitzing resistance, $n$ is the particle density,  $c_0$ is the interlayer separation for the cuprates- equalling half the c-axis lattice constant $c_L$ for the single layer compounds considered here. Here $\bar{\rho}$ is the dimensionless resistivity computed in terms of the microscopic model parameters and temperature measured in units of $t$. We express $\bar{\rho}$ in terms of the band  velocities $\vec{v}_{\vec{k}}=\vec{\partial}_{\vec{k}} \varepsilon_k$, the Fermi function $f(\omega)=\{e^{\beta \omega}+1\}^{-1}$  and the electron spectral  function $A(\vec{k},\omega)$ obtained from the ECFL formalism \cite{ECFL} as
\beq
\frac{1}{\bar{\rho}}= \frac{(2 \pi)^2}{a_0^2} \int_{-\infty}^\infty d\omega (-\frac{\partial f(\omega)}{\partial \omega}) \langle A^2(\vec{k},\omega) (\hbar v_k^x)^2 \rangle_{\vec{k}} . \label{convolute}
\eeq
 The  behaviour of $\bar{\rho}$ is quite intricate and it is discussed below as a function of various parameters. 
 
 For typical parameters encountered in our study, the resistivity $\rho$  is found to be linear in T in a certain range of temperature, wherein one can express it in a Drude type form  $\rho=\frac{m_{*}}{n_{*} e^2 \tau} $, where the relaxation time  ${\tau}= \frac{h}{k_B T}$ involves only  Planck's constant. This is sometimes referred to as the ``Planckian limit'' \cite{Cooper2}, which is free from any material specific scale. Taking this observation as seriously suggesting a universal and otherwise scale free physics seems hard.  
 It is impossible to extract   $\tau$ from  experiments- unencumbered by other essential parameter such as $n_{*},m_{*}$.
 The parameters $n_{*},m_{*}$ in such a fit can be determined in each case and  are far from being invariant - they vary with all other basic parameters of the theory. A similarly non-universal situation seems to occur in most experiments as well, where  specific sets of data show a linear in T behaviour over a restricted range.

\subsection{Computation} 

\textcolor{black}{For Tl2201 and Bi2201} the ECFL equations were solved iteratively on four $N_k\times N_k$ lattices with $N_k$ = 81, 86, 91 and 96, with a frequency grid of $N_\omega=2^{14}$ points. In \cite{Aspects-I}  smaller systems  $N_k$=62 were studied, but otherwise we used the identical computational procedure. \textcolor{black}{Even at these larger sizes} our systems are still too small to display the systematics expected from  finite-sized scaling analysis. The different sizes studied show small but unsystematic variations. These are treated by averaging the resistivity results over the four samples. With a few exceptions at the lowest T values fluctuation $\delta \rho/\rho$  is generally less than 2\%. \textcolor{black}{Hg1201 was solved at  $N_\omega=2^{12}$ and $N_k=92$ which we have estimated to be sufficiently accurate given the significantly smaller $t$ value.}

 Also as in \cite{Aspects-I}  theoretical resistivities extending below $T/t$=77.8 K/eV are found by extrapolating from a fit $\rho\sim \alpha \frac{T^2}{T+T_0}$.

\section{Bi2201 Results}
\subsection{Fermi surface and band parameters  of Bi2201} 

\begin{figure}[H]
\centering
\includegraphics[width=.49\textwidth]{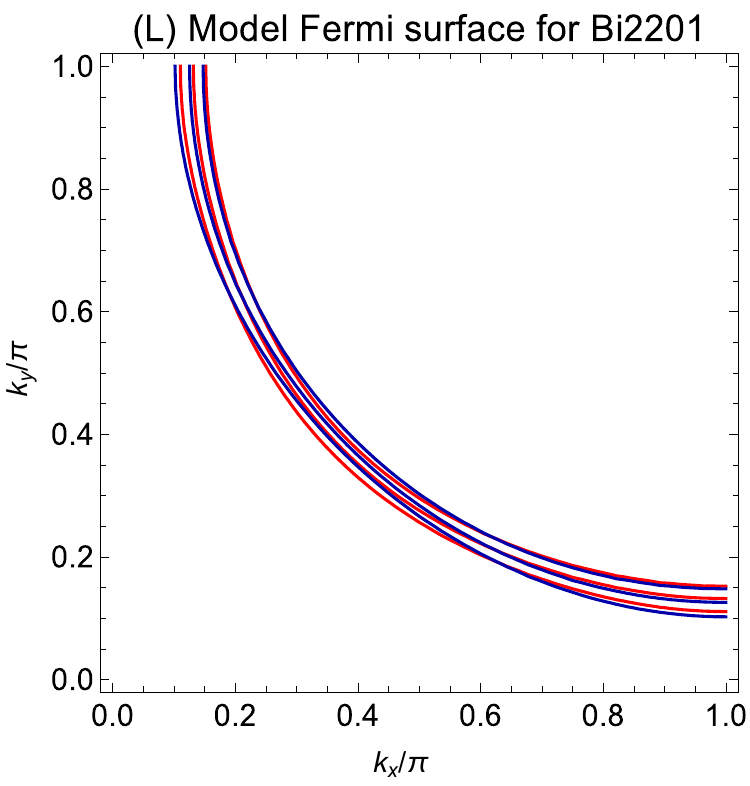}
\includegraphics[width=.49\textwidth]{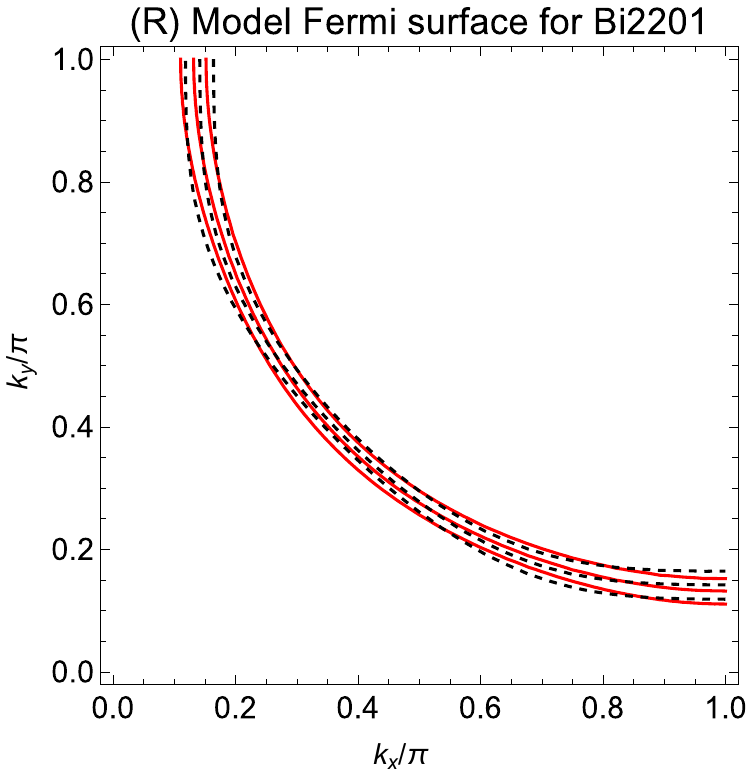}
\captionof{figure}{ The band parameters used here  are  given in \disp{tb-Bi2201}. The resulting Fermi surfaces at  densities $n=0.75,0.8,0.85$ are  shown in red in the two figures and compared to  those from other suggested models,   ({\bf Left})   with  parameters $t'/t=-0.3143$, $t''/t=0.04286$ (in blue) quoted in \cite{Hussey} and, ({\bf Right}) with  parameters $t'/t=-0.156$, $t''/t=0.164$ (dashed lines) quoted in \cite{Bansil}.}
\label{Bi2201-BS}
\end{figure}
We study Bi2201 using the tight binding parameters
\beq
\mbox{\em Bi2201 tight binding parameters: }&&t'=-0.4 t,t''=0.0,J=0.17 t , \nonumber \\
&&t= 1.176  eV \label{tb-Bi2201}\nn \\
&&c_L,c_0=24.6,12.3 \AA \phantom{.}\cite{Hussey}
\eeq
where the  magnitude of  $t$ is estimated from a best fit with the resistivity over all available samples, as discussed below  in \figdisp{Bi2201-Fig4}.  Our choice  in \disp{tb-Bi2201} is guided by requiring the simplest parameterization, with the smallest number of non-zero hopping elements, and differ somewhat from other schemes in literature. The band parameters suggested in \cite{Hussey}, upon conversion to the convention used here  are 
expressible in the form $t'=-0.3143\, t, t''=0.04286 \, t$, and earlier estimates from band theory \cite{Bansil} are farther away $t'=-0.156 \, t, t''= 0.164 \, t$. \figdisp{Bi2201-BS} shows  that both of these alternate schemes lead to very similar Fermi surfaces found from \disp{tb-Bi2201} . While $t'/t$ and $t''/t$ are obtainable from the measured Fermi surface when available, the magnitude of $t$ remains undetermined by these considerations. The magnitude of the single theoretical parameter $t$ is determined  to give a good overall fit to the resistivities over available densities, as noted in \figdisp{Bi2201-Fig4}. We also made   a few further checks with the the parameterization in \cite{Hussey}, which yielded very similar resistivities after adjusting the scale of $t$.

We first summarize the available samples from \cite{Hussey,Martin,Fiory} in \tabdisp{Tab-Presland}, and discuss their resistivity in detail below.  Their $T_c$'s and other parameters  are listed in \tabdisp{Tab-Presland}. In the last row of \tabdisp{Tab-Presland} we also include the early measurement of \cite{Martin,Fiory}. Here  we review those early findings in the context of recent and modern measurements in \cite{Hussey}, as well as calculations from the ECFL theory.

\begin{table}[h]
\centering
\begin{tabular}{ | p{0.6 in}| p{ .6in} |p{0.6 in }| p {.7 in}| p{0.6 in}| }\hline 
Sample \#&Ref. &$T_c$ in K & x (\disp{Presland}) & $T_{max}$ in K \\ \hline
S:1&\cite{Hussey}&7&0.258&300\\ \hline
S:2&  \cite{Hussey}&17&0.239&300\\ \hline
 S:3& \cite{Hussey}&27&0.213&300
  \\ \hline
  S:4&\cite{Hussey}&31&0.197&300\\ \hline \hline
  S:5&\cite{Martin}&6.5&0.259 (?)&800\\ \hline
\end{tabular} 
\caption{Samples S:1-S:4 of Bi2201 studied in 2022\cite{Hussey}, and sample S:5  studied in  1989\cite{Martin,Fiory} are   compared with theory below. Resistivity measurements are reported up to $T_{max}$. In \cite{Hussey} the observed $T_c$ for each of these overdoped samples is used to estimate the hole density  $x$ using the phenomenological relation \disp{Presland}.  The quoted $T_c$ of sample S:5 \cite{Martin,Fiory}  converts to a density $x$=0.259  by using \disp{Presland}. This value is  essentially  identical to that of sample S:1 in \cite{Hussey}, but is observed to have a substantially different magnitude of resistivity from it, as seen in \figdisp{Bi2201-Fig3}.
Theoretically (see \figdisp{Bi2201-Fig2}) $x$=0.32  seems overall to be  more consistent for sample S:5.}
\label{Tab-Presland}
\end{table}

In \cite{Hussey} the normal state resistivity of samples  S:1-S:4 are reported for temperatures up to 300K.
The  question of determining the hole density $x$ in this system is discussed in  \cite{Hussey}. They estimate  $x$ ( $=p$ ) by comparing the observed resistivity $\rho(T)$ and $d \rho(T)/dT$ with observations on   LSCO at different densities \cite{Ando-1}. They observe that for $x$ deduced from different arguments, such as the ARPES Luttinger count, comparing  resistivity and its T derivatives and the phenomenological relation (\disp{Presland}) between $T_c$ and $x$   lead to rather different results in general. 
For the samples studied further in this work,  we could not find the recommended estimates of $x$ in  the paper \cite{Hussey}, and therefore
 used \disp{Presland} to arrive at the $x$- column using the quoted $T_c$ values, as detailed in \cite{expl-x}. Since the ECFL calculation- with suitable parameters- leads to a consistent quantitative description of the LSCO resistivity $\rho$ and the derivative $d \rho/dT$  data from \cite{Ando-1}, we go ahead and compare the current calculation with all the reported resistivity data below- where we evaluate and comment on the quoted $x$ values as well as make suggestions to revise them.  

\subsection{Resistivity of Bi2201} 

\begin{figure}
\centering
\includegraphics[width=.47\textwidth]{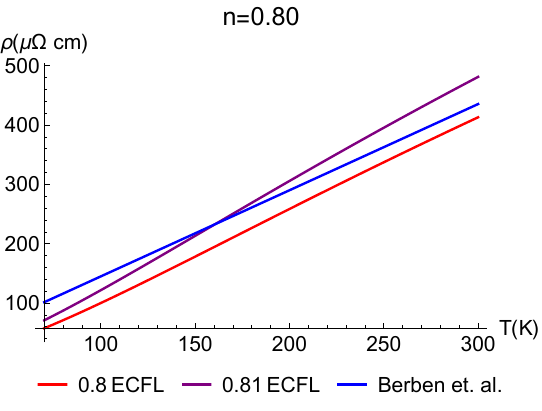}
\includegraphics[width=.47\textwidth]{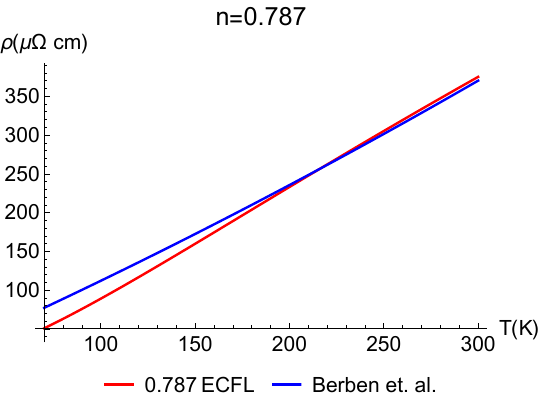}\\
\includegraphics[width=.49\textwidth]{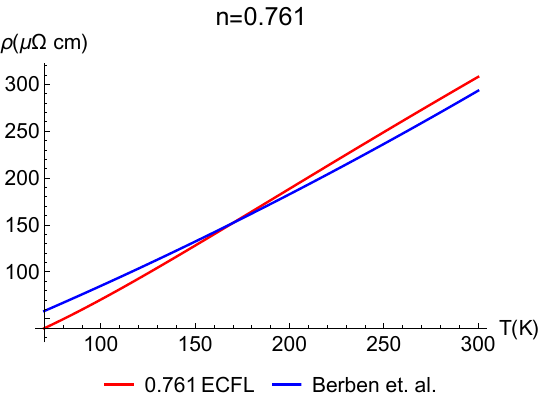}
\includegraphics[width=.49\textwidth]{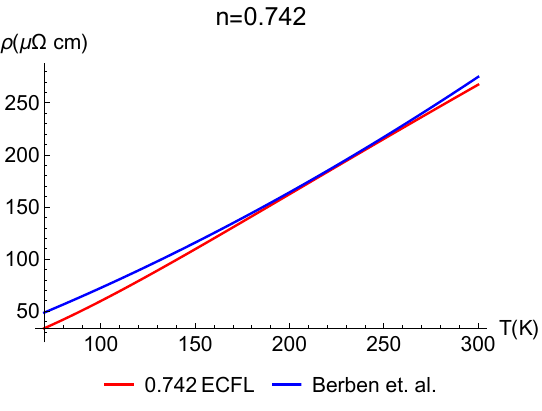}\\
\captionof{figure}{\textcolor{black}{Bi2201 comparisons.} From top left-the ECFL resistivities (in red and magenta) at $n=0.80$ and $0.81$,  then (in red) at $n=0.787,0.761,0.742$  plotted against the resistivity data  of Berben {\em et. al.} \cite{Hussey} for samples S:4,S:3,S:2,S:1 respectively. The experimental data has been  adjusted for impurity contribution by a simple shift in each case.The top left panel shows the theoretical  ECFL resistivities at $n=0.81$ ($x$=0.19) as well as $n=0.8$ ($x$=0.2), which seems to bracket the data.  The ECFL curves use the band parameters in \disp{tb-Bi2201} with $t=1.176$ eV  for all the curves. This value of $t$ seems to be reasonable for the overall available data set, and the tight binding parameters \disp{tb-Bi2201} are used in calculation of all the Bi2201 figures below.  For  T\lessim 70K the ECFL results are parabolic in T. 
}
\label{Bi2201-Fig4}
\end{figure}
In \figdisp{Bi2201-Fig4} we compare the ECFL theory resistivity with that from samples S:1-S:4 of \cite{Hussey}. We note that the $d\rho/dT$ of the two sets are close, however the sample S:4 has somewhat bigger $\rho$ than the theoretical estimate-indicating that the  estimated $x$ might be slightly off. For this purpose, the top left panel in \figdisp{Bi2201-Fig4} shows the ECFL resistivities at $n=0.81$ ($x$=0.19) as well as $n=0.8$ ($x$=0.2) with identical remaining  parameters, which seem to bracket the experimental result for the sample S:4.

\begin{figure}
\centering
\includegraphics[width=.7\textwidth]{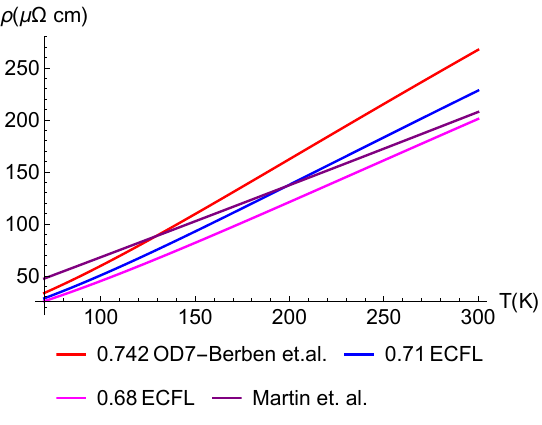}
\captionof{figure}{\textcolor{black}{Bi2201 comparisons.} ECFL resistivity at $n=0.68$ and $n=0.71$  plotted against the data S:5 from Martin. et. al. \cite{Martin,Fiory}  \textcolor{red}and S:1 from Berben et. al. \cite{Hussey}.}
\label{Bi2201-Fig3}
\end{figure}

\begin{figure}
\centering
\includegraphics[width=.8\textwidth]{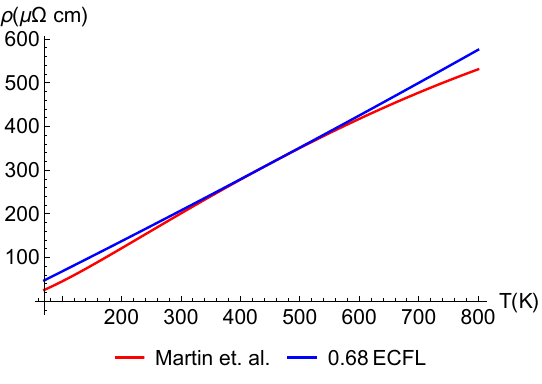}
\captionof{figure}{\textcolor{black}{Bi2201 comparisons.} ECFL resistivity at $n=0.68$ (blue) plotted against the data from Martin. et. al. \cite{Martin,Fiory} over a wide T range.}
\label{Bi2201-Fig2}
\end{figure}

Turning to the data from S:5 \cite{Martin,Fiory}, in \figdisp{Bi2201-Fig3} we compare the resistivity with S:1 from \cite{Hussey}.
In \tabdisp{Tab-Presland}, we see that by using the phenomenological relation \cite{Pres,expl-x}, these two are expected to be very close, but the resistivities do not appear to be very close. We next compare these with  
the ECFL resistivities at $n=0.71,0.68$ using the previously determined value  $t$=$1.176$eV. It seems thus that these two curves bracket the result for S:5. We explore this further by plotting the resistivity over a much bigger T scale- up to 800K in \figdisp{Bi2201-Fig2}. It is seen here that there is reasonable match between the two curves over most of the range.

\begin{figure}
\centering
\includegraphics[width=.8\textwidth]{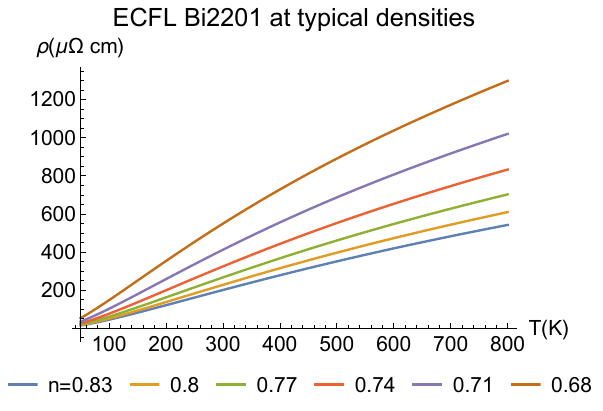}
\captionof{figure}{A summary of ECFL resistivities at typical densities over a wide temperature window.  }
\label{Bi2201-All}
\end{figure}

In \figdisp{Bi2201-All} we display the ECFL resistivities using the band parameters in \disp{tb-Bi2201} over a wide set of densities and a broad range of T. We note that the nonlinear (usually quadratic) corrections to the resistivity  become more evident as the particle density $n$ increases, being almost linear over the whole range at the lowest density- as also seen in \figdisp{Bi2201-Fig2}. 

\begin{figure}[t]
\centering
\includegraphics[width=.6\textwidth]{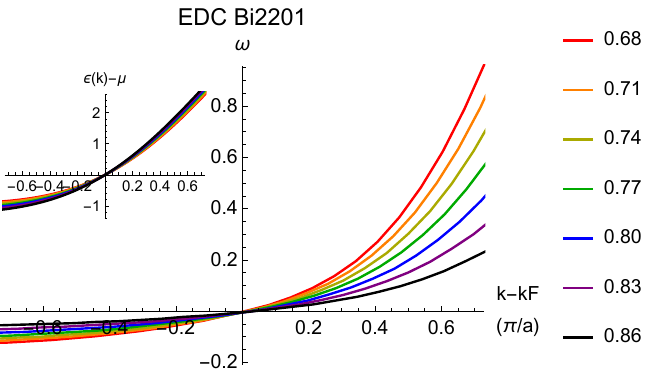}
\captionof{figure}{  \sri{Bi2201 energy distribution  curve (EDC) dispersions across the full range of densities. The inset displays the bare band dispersion minus the chemical potential, $\epsilon(k)$-$\mu$ in units of $t$. The ratio of their slopes gives an estimate of the effective masses, which are listed in Table \ref{Table-2}}}
\label{EDC_Bi2201}
\end{figure}

\textcolor{black}{We also include energy distribution curve (EDC) dispersions sampled in the nodal $k$ direction at our lowest available temperature (91.5K) in \figdisp{EDC_Bi2201}. EDCs are slices of the spectral function across frequency at fixed $k$. The dispersion shows the $\omega$ value at which the spectral peak is found in each $k$ slice plotted against $k-k_F$. A comparison of the dispersion with the band dispersion for the same parameters as shown in the inset gives an estimate of the effective band mass $\frac{m^*}{m}$. See \tabdisp{EDCBi} for our effective mass calculations.}

\begin{table*}[h]
\begin{center}
\begin{tabular}{ |c|c|c|c|c|c|c|c| } \multicolumn{8}{c}{{\bf  Bi2201}}  \\ 
 \hline
 Density & 0.68&0.71&0.74&0.77&0.80&0.83&0.86 \\ \hline
 $m^*/m$&7.460&8.462&9.345&10.782&13.291&18.251&23.795\\\hline 
\end{tabular}
\end{center}
\caption{\textcolor{black}{Bi2201  $m^*/m$ defined as  the ratio of  slopes of $\epsilon(k)$  and the EDC dispersion at $k=k_F$, shown in \figdisp{EDC_Bi2201}. \label{Table-2}}}
\label{EDCBi}
\end{table*}

\section{Tl2201 Results}

\subsection{Fermi surface and band parameters  of Tl2201} 

\begin{figure}[H]
\centering
\includegraphics[width=.49\textwidth]{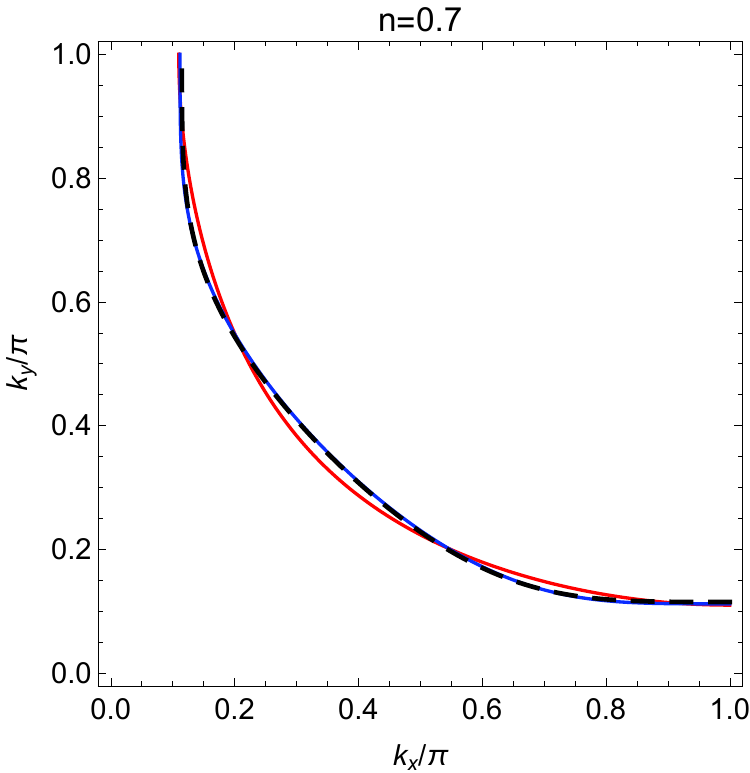}
\includegraphics[width=.49\textwidth]{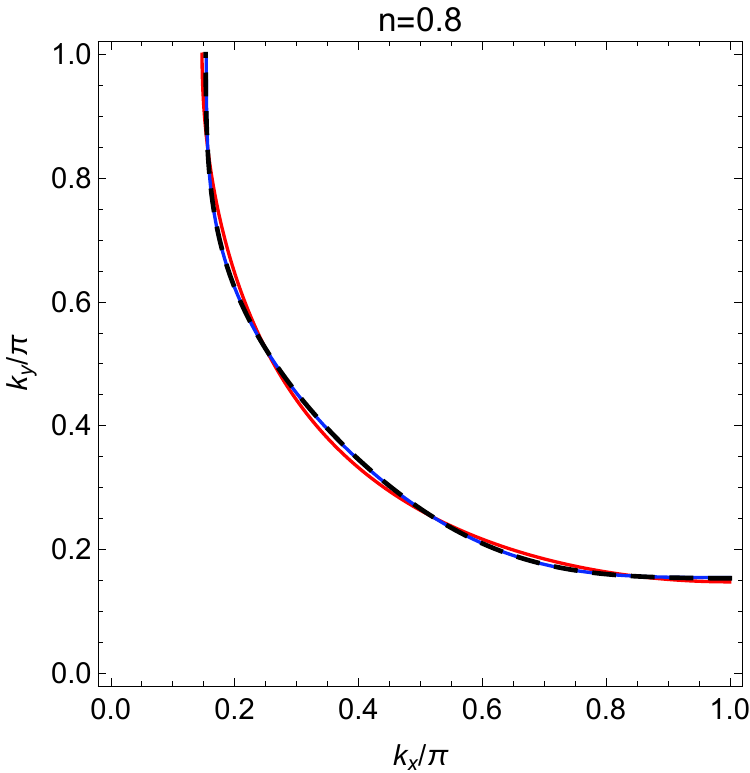}
\captionof{figure}{(Left) The Fermi surfaces at $n$=0.7. The blue curve is from Model-B (\disp{tb-Tl2201})
and the red curve is from Model-A (\disp{tb-Tl2201}). \textcolor{black}{The black dashed line shows the Fermi surface from the experimentally derived energy dispersion in  \disp{bandexp}.} (Right) the same curves at density $n$=0.8. We thus see that the result from Model-A is fairly close to the experimentally derived Fermi surface, while the Model-B is nearly exact at these densities.}
\label{Tl2201-FS}
\end{figure}

The ARPES  determined Fermi surface for Tl2201 is available in \cite{Tl2201,Cooper1}. This work  fits it to a   band structure 
\begin{multline}
    \epsilon(k_x,k_y)=\frac{1}{2}\tau_1(\cos(k_x)+\cos(k_y))+\tau_2\cos(k_x)\cos(k_y)+\frac{1}{2}\tau_3(\cos(2k_x)+\cos(2k_y)) \\
    +\frac{1}{2}\tau_4(\cos(2k_x)\cos(k_y)+\cos(k_y)\cos(2k_y))+\tau_5\cos(2k_x)\cos(2k_y)
    \label{bandexp}
\end{multline}
where in units of eV $\tau_1=-0.725$, $\tau_2=0.302$, $\tau_3=0.0159$, $\tau_4=-0.0805$ and $\tau_5=0.0034$.
Our preference is to use fewer parameters for performing the ECFL calculations involving 
a large number of further steps. Hence we checked for the possibility of fitting the Fermi surface resulting from \disp{bandexp} with at most two sets of neighbours i.e. with $t,t',t''$ only,  and found that there are two distinct type of parameters which provide excellent fits of the above Fermi surface over the full range of densities studied- as seen in \figdisp{Tl2201-FS}. We refer to these as Model-A and Model-B. The two hopping variable sets are given by
\beq
\mbox{\em Tl2201 tight binding parameters: }& \mbox{Model-A}&t'=-0.430 t, t''=0.005 t  \nonumber \\
&&t= 1.82  eV, J=0.17 t , \label{tb-Tl2201}\nonumber \\
&\mbox{Model-B}&t'=-0.237 t, t''=0.138 t \nonumber \\
&& t=1.053  eV, J=0.17 t \label{tb-Tl2201}\nn \\
&c_L,c_0&=23.1, 11.56 \AA \phantom{.} \cite{Cooper1} \label{tb-Tl2201}
\eeq
and we included the standard value of $J$ used for easy reference.

 We  display in \figdisp{Tl2201-FS} the Fermi surfaces from \disp{tb-Tl2201} compared with the Fermi surface from \disp{bandexp}.

\subsection{Resistivity of Tl2201} 

\begin{figure}
\centering
\includegraphics[width=.63\textwidth]{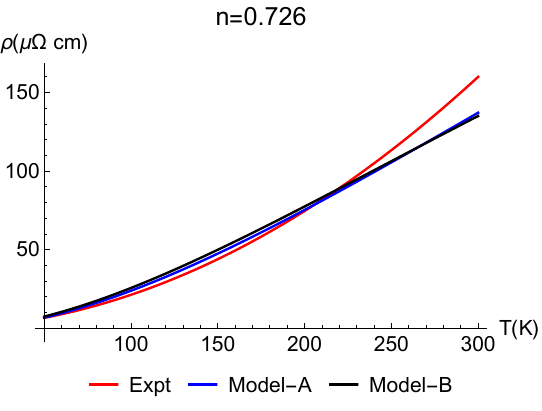}
\captionof{figure}{ECFL resistivity for Tl2201 using parameters in \disp{tb-Tl2201} at a density $n$=0.726 with Model-A (blue) and Model-B (black), compared with the experimental curve from \cite{Cooper2}. The two values of $t$ for the two models quoted in \disp{tb-Tl2201} are fixed by fitting the theoretical  temperature with the observed one, and are taken to be fixed for other densities. We see that the theoretical curves as well as the experimental one show a significant quadratic correction in T here and at most other densities.  }
\label{Tl2201-726}
\end{figure}

\begin{figure}
\centering
\includegraphics[width=.63\textwidth]{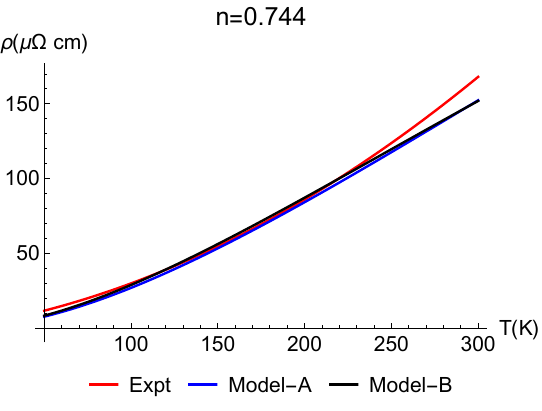}
\captionof{figure}{ECFL resistivity for Tl2201 using parameters in \disp{tb-Tl2201} at  a density $n$=0.744 with Model-A (blue) and Model-B (black), compared with the experimental curve from \cite{Cooper2}.
Below 250 K, the theoretical and experimental curves are seen to be close  at this density.
}
\label{Tl2201-744}
\end{figure}

\begin{figure}
\centering
\includegraphics[width=.63\textwidth]{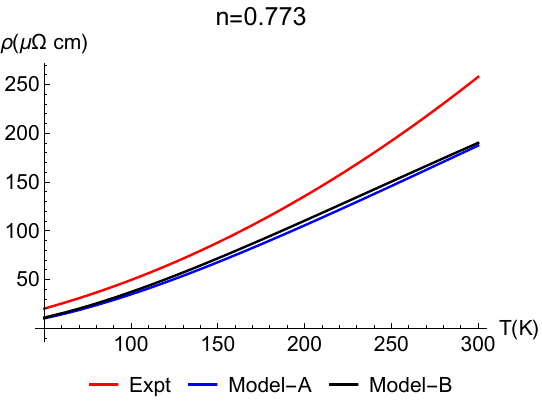}
\captionof{figure}{ECFL resistivity for Tl2201 using parameters in \disp{tb-Tl2201} at a density $n$=0.773 with Model-A (blue) and Model-B (black), compared with the experimental curve from \cite{Cooper2}.  The experimental  curve is  somewhat shifted upwards from the theoretical one. 
}
\label{Tl2201-773}
\end{figure}

\begin{figure}
\centering
\includegraphics[width=.63\textwidth]{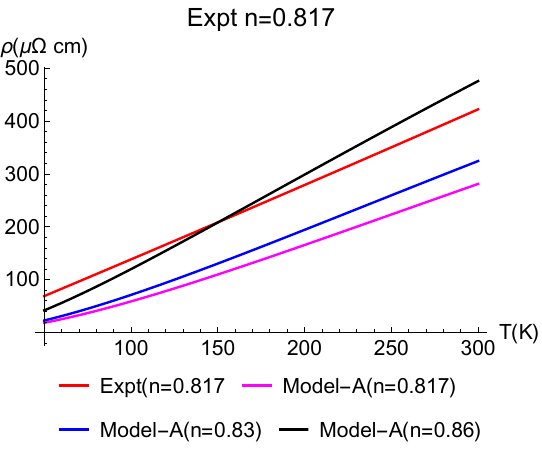}
\captionof{figure}{ECFL resistivity for Tl2201 using parameters in \disp{tb-Tl2201} at a density $n$=0.83 (blue) and 0.86 (black) compared with the experimental curve at $n$=0.817. The two theoretical curves bracket the experimental curve, while the theoretical curve at $n$=0.817 is noticeably below the data.}
\label{Tl2201-817}
\end{figure}

Ref.~\cite{Cooper2} presents the normal state resistivity of four samples with densities n=$0.817,0.773,0.744,0.726$.
In \figdisp{Tl2201-726} - \figdisp{Tl2201-773} we compare \sri{ theoretical resistivities from ECFL for } model A and model B for $n=$ 0.726, 0.744 and 0.773 to experimental results from Cooper et. al. \cite{Cooper2}. In general the resistivities of Model-A and Model-B are very close over all densities and temperatures. \figdisp{Tl2201-817} shows experimental results for $n=0.817$. This curve does not agree well with either of  our models and seems to be somewhat higher in magnitude. Two higher density results for model A are displayed for additional comparison, the curve at $n=0.86$ seems  closer in scale to the data. Further data at nearby densities would be helpful to clarify the resistivity-density systematics. 

\begin{figure}[H]
\centering
\includegraphics[width=.8\textwidth]{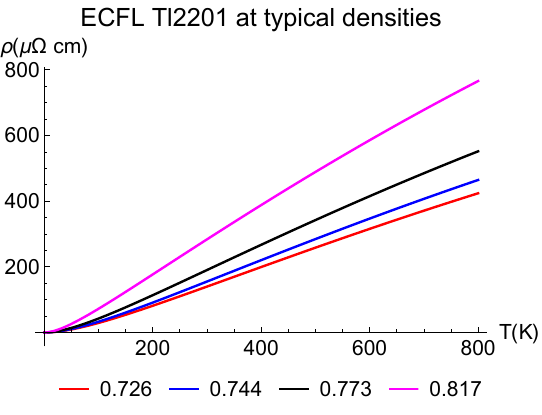}
\captionof{figure}{ECFL resistivities \sri{for Tl2201} at typical densities over a larger temperature window using Model-B.  }
\label{Tl2201-All}
\end{figure}

In \figdisp{Tl2201-All}  we  display the full set of results at different densities for Model B over a wide range of T, Model-A gives very similar results and is therefore not displayed.

\begin{figure}[H]
\centering
\includegraphics[width=.49\textwidth]{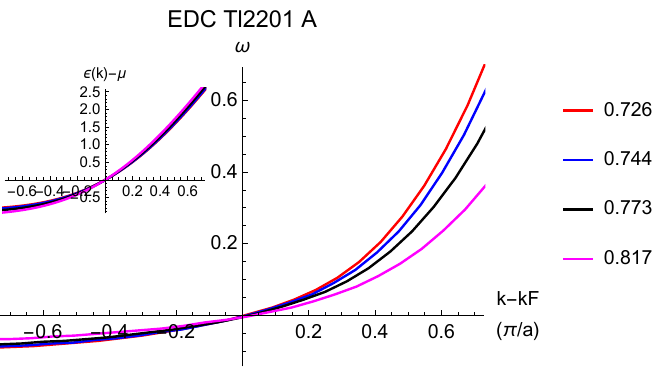}
\includegraphics[width=.49\textwidth]{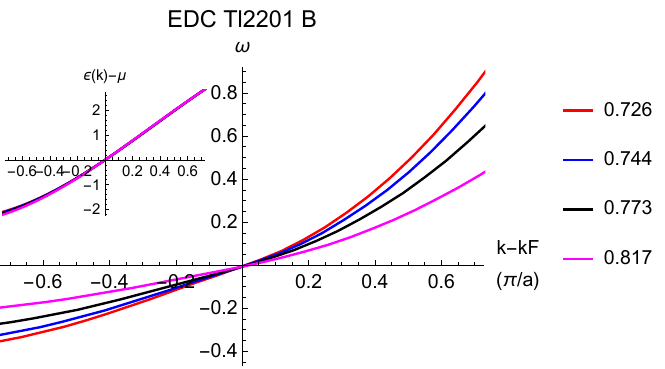}
\captionof{figure}{\sri{Tl2201 Model A and Model B energy distribution  curve (EDC) dispersions across all four densities. The inset displays the bare band structure minus the chemical potential, $\epsilon(k)$-$\mu$ in units of $t$. The corresponding effective masses are given in Table \ref{Table-3}.} \label{Figure-13}}
\label{EDC_Tl2201}
\end{figure}

\textcolor{black}{We also include EDC dispersions sampled in the nodal direction at our lowest available temperatures (141.6K for Model A, 81.9K for Model B) in \figdisp{EDC_Tl2201} for both of our sets of band parameters. Comparison of the dispersion with the band dispersion gives an estimate of the effective band mass $\frac{m^*}{m}$ as shown in \tabdisp{EDCTl}. The lowest density, 0.726, has an effective mass of 8.056 in Model A and 6.802 in model B. This density is cited in \cite{Cooper2} as having the lowest $T_c$ of the set, 26.5 K. In \cite{Vignolle} an overdoped Tl2201 sample with a lower $T_c$ (15 K) is referenced as having $m^*/m=4.1\pm 1$.  Overall,  this seems to fit with the pattern observed in our $m^*/m$ calculations, with $m^*/m$ lowering as $n$ (indicated by $T_c$) decreases.}

\begin{table*}[h]
\begin{center}
\begin{tabular}{ |c|c|c| } 
\multicolumn{3}{c}{{\bf  Tl2201}}  \\ 
 \hline
 Density & Model A $m^*/m$  & Model B $m^*/m$\\ \hline\hline
 0.726 & 8.056 & 6.802 \\ \hline
 0.744 & 9.386 &  7.259 \\ \hline
 0.773 & 10.854 &  8.351 \\ \hline
 0.817 & 15.443 & 11.961 \\
 \hline
\end{tabular}
\end{center}
\caption{\textcolor{black}{Tl2201 $m^*/m$ defined as  the ratio of  slopes of $\epsilon(k)$  and the EDC dispersion at $k=k_F$ shown in \figdisp{Figure-13}} \label{Table-3}}
\label{EDCTl}
\end{table*}

\section{Hg1201 Results}

\subsection{Fermi surface and band parameters  of Hg1201} 

\textcolor{black}{ARPES results in \cite{Vishik} provide a set of tight binding parameters for Hg1201 as shown in \disp{tb-Hg1201}}. 

\beq
\mbox{\em Hg1201 tight binding parameters: }&&t'=-0.228 t,t''=0.174t,J=0.17 t , \nonumber \\
&&t= 0.2  eV \label{tb-Hg1201}\nn \\
&&c_L,c_0=19, 9.5 \AA \phantom{.}\cite{Yamamoto2}
\eeq

\subsection{Resistivity of Hg1201} 

\begin{figure}[H]
\centering
\includegraphics[width=.49\textwidth]{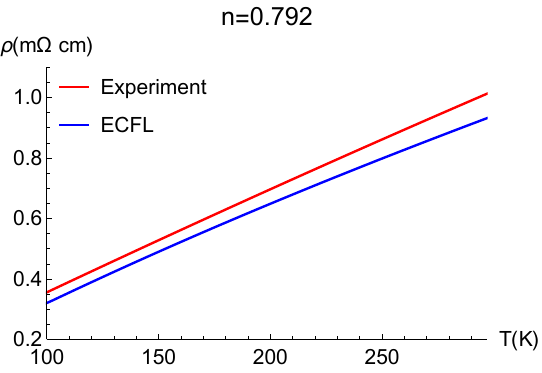}
\includegraphics[width=.49\textwidth]{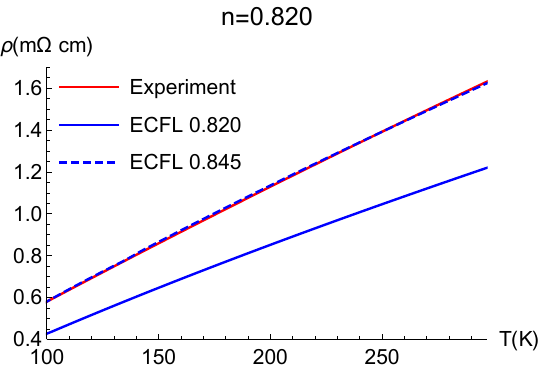}
\includegraphics[width=.49\textwidth]{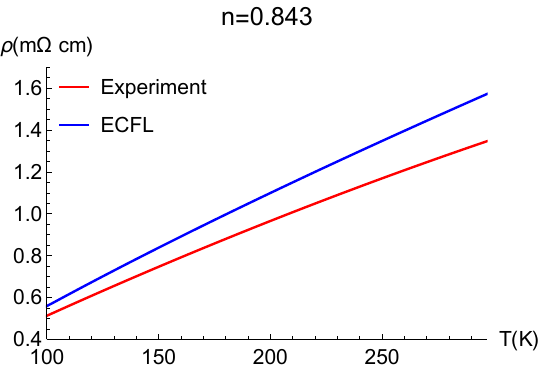}
\includegraphics[width=.49\textwidth]{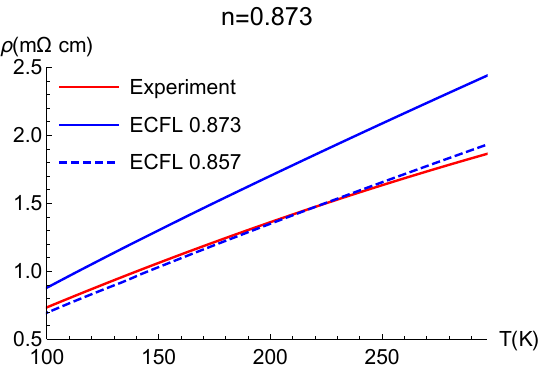}
\captionof{figure}{\textcolor{black}{Hg1201 comparisons: ECFL resistivities (blue) compared to resistivity data from \cite{Yamamoto} fit down to T=0 with impurity resistivity subtracted. For $n=$ 0.820 and 0.873 a second ECFL curve is also displayed in a dashed line, showing a different density that more closely matches the experiment.}}
\label{Hg_exp}
\end{figure}

\textcolor{black}{Resistivity data for this system is available from Refs.\cite{Yamamoto,Greven}. Since the ECFL theory has been developed and tested in the optimum to overdoped regimes, we focus on data within this regime. Ref. \cite{Yamamoto} provides resistivity for samples with a wide range of explicitly stated dopings. Amongst these  we focus on the four resistivities spanning $n=0.792-0.873$,
in the optimum to overdoped regimes. This range overlaps with the range studied in our other materials. Data was taken from plots in \cite{Yamamoto} using the data extraction tool DigitizeIt.  From the data  an  impurity resistivity contribution was estimated and subtracted off for our comparison to theory.  Our comparisons between ECFL and experiment are found in \figdisp{Hg_exp}.}

\sri{In common with Bi2201, the electronic density in the Hg1201 material is reported too difficult to assign, as compared to other single layer cuprates. Keeping this in mind, we display in \figdisp{Hg_exp} the theoretical results at densities  quoted in the corresponding experiments \cite{Yamamoto}, and for $n=0.820$ and $n=0.873$, at one additional density, somewhat different  from the nominal one, where data matches theory somewhat better. While this empirical procedure is suggestive of the origin of the discrepancies, it is not entirely satisfactory from a theoretical perspective.  }

\begin{figure}[H]
\centering
\includegraphics[width=.8\textwidth]{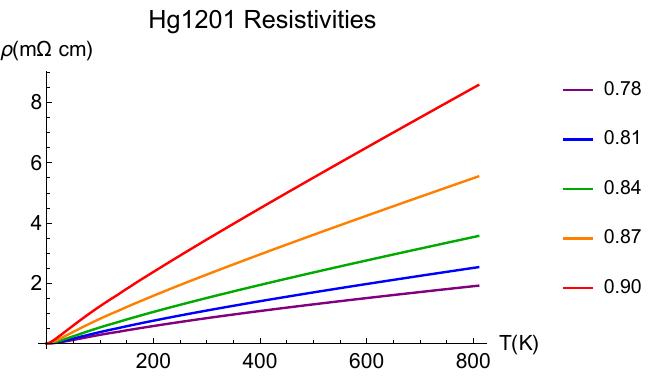}
\captionof{figure}{\textcolor{black}{The full set of ECFL resistivities for Hg1201. Other densities used below are found from an interpolation of these.}}
\label{Hg_all}
\end{figure}

The  full results of ECFL can be seen in \figdisp{Hg_all}

\section{Discussion of Results}

We  first comment about a minor difference in the treatment of the impurity contribution to resistivity in this work from that in \cite{Aspects-I}. In the case of Tl2201, the data for each sample  presented in \cite{Cooper1,Cooper2} is in the convenient form of a fit to a simple function $\rho= \rho_0+\rho_1 T + \rho_2 T^2$, and hence we drop the term with $\rho_0$ to compare with theory.
We note that for the case of Bi2201 and Hg1201,  we digitized the published data and fit it to a convenient functional form, and followed the same recipe.

Our results for the single layer compound Bi2201 are compared with theory in \figdisp{Bi2201-Fig4}. Theory is in reasonable accord {\em on an absolute scale} with the data from  \cite{Hussey} at $n$=0.787, 0.761 and 0.742. At $n$=0.80 the theoretical result for $n$=0.80 is somewhat off from the data, while the result for $n$=0.81 is close- albeit with a slightly greater slope. There seems to be no single scaling of t which could improve matters at all densities. A notable aspect of the comparison is that the data  as well as theory show a $T^2$ correction to linear behaviour of different extent depending on the density.

The density of the sample in \cite{Martin} was not fixed precisely, as far as we could see. With optimism that might be questionable,  we  estimated it crudely from the observed $T_c$, using the phenomenological relation \disp{Presland} to be $n$=0.74. This estimate roughly coincides with the density of sample S:1 of \cite{Hussey}. In \figdisp{Bi2201-Fig3} the data at $n$=0.742 from \cite{Hussey} and the data from \cite{Martin} are compared, together with the theoretical curves from ECFL at densities $n$=0.71 and $n$=0.68.  The theoretical curves are drawn assuming the parameters already determined from the data sets from \cite{Hussey}.
These densities are somewhat lower than the theoretical curve at $n$=0.742 shown in \figdisp{Bi2201-Fig4}, but seem to bracket the data of \cite{Martin}, suggesting that for some unclear reason, the density of the sample in \cite{Martin} is close to $n$=0.68.   We take this phenomenological possibility further in \figdisp{Bi2201-Fig2} where the theoretical curve at $n$=0.68 and the data from \cite{Martin} are compared. Barring the limiting values of T, the match between theory and the data seems intriguing, especially given the broad range of temperatures - up to 800 K.

Turning to Tl2201, 
in \figdisp{Tl2201-726,Tl2201-744,Tl2201-773,Tl2201-817} we compare the data at densities $n$=0.726, 0.744, 0.773 and 0.817 with theoretical results found using  the two band models described in \disp{tb-Tl2201}.  The two theoretical models, start from two rather different sets of parameters characterized by distinct $t'/t,t''/t$ values, and somewhat surprisingly describe the  Fermi surface shape almost equally well, as seen in \figdisp{Tl2201-FS}. It is therefore of interest to note that the resistivities of the two models agree very well, after a suitable choice is made of the nearest neighbour hopping $t$ for each model, and seems to confirm the initial belief that the Fermi surface shape largely determines the resistivity results.  We note that the data for $n$=0.726 and $n$=0.744 agrees on an absolute scale with theory, whereas at a higher densities $n$=0.773 the data is parallel but offset from the theoretical curves. At $n$=0.817 the discrepancy between theory and experiment is greater than at lower densities. To quantify this, we also display the calculated resistivity at $n$=0.83 and $n$=0.86 along with $n$=0.817. It is interesting that the theoretical curve for $n$=0.86 has the same scale as the experiment, and it might be interesting to  obtain data from samples with other densities in this range. 

\textcolor{black}{For Hg1201 we have calculated ECFL results for a range of densities as shown in \figdisp{Hg_all}, and we have made interpolations from these curves for comparison to four of the densities ($n=$ 0.792, 0.82, 0.843 and 0.873) found in \cite{Yamamoto}, as seen in \figdisp{Hg_exp}. The data for $n$=0.792 and $n$=0.843 are reasonably close to the theoretical curves, while $n$=0.82 and $n$=0.873  theory and data are parallel over this range, but display a vertical shift. For illustrative purposes we provide alternate densities from the nominal (quoted) ones where the matching is closer. }

\sri{It might be useful to visualize the full set of theoretical resistivities, their systematic progression with density and temperature. For this purpose,
we display the theoretical resistivities  for Bi2201  in \figdisp{Bi2201-All}  over a broad range of temperatures for six  densities. Similarly  theoretical resistivities for Tl2201 {Model B} are provided in   \figdisp{Tl2201-All}, and  for Hg1201 in \figdisp{Hg_all}}.

\section{ Concluding remarks}

\sri{In the present and our earlier work \cite{Aspects-I} we have performed a detailed application of the extremely correlated Fermi liquid theory to calculate the normal state resistivity of  single layer High $T_c$ cuprate materials in the optimum to overdoped density regimes. The relative simplicity of the single layer materials compared to other strongly correlated  materials arises from the almost decoupled nature of the layers, so that a purely 2-dimensional description, ignoring motion in 
the third direction, is quite reasonable. The ECFL theory requires a very few parameters- detailed in \disp{r-formula}- and yields resistivity on an absolute scale. Comparing the results with experiments is therefore feasible. It becomes especially meaningful, provided a large enough set of materials and data sets are included. It is of interest to see if the wide variety of experimentally seen behavior- with variable T dependence and non-trivial  density dependence-  can be reproduced quantitatively  by the theory.  
}

\begin{table}[H]
\centering
\begin{tabular}{||p{3.2cm} |c|c|c|c|c||}
\multicolumn{6}{c}{{\bf  All Single Layer Cuprate High $T_c$ Materials}}  \\ \hline \hline
{\bf  Hole-doped } &x-range ($N_{samp}$)&$T_{max}$(K)&$t'/t$& $t''/t$  & t (eV)   \\ \hline \hline
 $La_{2-x}Sr_xCuO_4$ (LSCO)\; \cite{Ando-1} &0.12-0.22 (11)&400& -0.2  &0& 0.9   \\ \hline
 $Bi_{2}Sr_{2-x}La_xCuO_6$  (BSLCO) \; \cite{Ando-1}&0.12-0.18 (7)&300& -0.25  & 0&1.35  \\ \hline 
 $Bi_2Sr_2CuO_{6+x}$ (Bi2201)  \cite{Hussey} &0.213-0.258 (4)&300&  -0.4   & 0 & 1.176 \\  
(Bi2201)  \cite{Fiory,Martin} &0.259\{0.32?\} (1)&800&-0.4&0&1.176    \\ \hline
 $Tl_2Sr_2CuO_{6+x}$ (Tl2201ModelA) \cite{Cooper2}&0.183-0.274 (4)&300& -0.430 & 0.005 &1.82  \\  
 (Tl2201ModelB) \cite{Cooper2}&&& -0.237 & 0.138 &1.053 \\ \hline 
 $HgBa_2CuO_{4+x}$ (Hg1201) \; \cite{Yamamoto}&0.127-0.208 (4)&300& -0.228 & 0.174 & 0.22  \\ \hline 
{  \bf Electron-doped } &&&&   &    \\ \hline \hline
 $Nd_{2-x}Ce_xCuO_4$ (NCCO) \; \cite{NCCO} &.125-.15 (2)&400& +0.2  &0& 0.9   \\ 
 $La_{2-x}Ce_xCuO_4$ (LCCO)\; \cite{LCCO}&.14-.17 (4)&300& +0.2 & 0&0.76   \\ \hline
 \hline 
\end{tabular}
 \caption{\textcolor{black}{For 37 samples belonging to 7 families of single layer cuprates, a comparison of experiments with the ECFL theory  is carried out  in this work and in \cite{Aspects-I}. The first five rows consist of the known hole doped single layer materials and the last two are the electron doped single layer materials. 
  We summarize the  range of hole density x (=1-$n$), the number of samples and  the temperature range studied in the first three columns.  In the last three columns we list  the band parameters used in the theory.}  }
\label{All-Compounds}
\end{table}

\sri{With the addition of results for the three  systems added here, to  the systems already studied in \cite{Aspects-I}, we now have a comparison between  all available single layer compounds and the   set of results of ECFL. In \tabdisp{All-Compounds} we display the full set of band parameters we deduced  to analyze these systems.

From a theoretical viewpoint,  the qualitative results for  the resistivity  $\rho$ in ECFL theory can be summarized as follows. We find  a $\rho \propto T^2$  behavior at very low $T \ll t/k_B$,  crossing over at a low T (still  at $T\ll t/k_B$) to  $\rho \propto T$  in  a ``strange metal'' regime. This behavior  is robustly realized in much of  the data. At a quantitative level, given the uncertainties in determination of material parameters, the comparison between theory and experiments found in our work  seems fair.  The originally surprising difference in the curvature observed between electron and hole doped resistivities, is also understandable within the theory, being related to a reversal of sign of the effective second neighbour hopping $t'/t$ \cite{Aspects-I}.

 Studies of  the $T$ dependence of the resistivity   have the potential of  uncovering the nature of the underlying many-body ground states in strongly correlated systems, i.e. to answer the question whether it is a  Fermi-liquid state of some sort or a non Fermi-liquid state.  The resistivity is inferred from the electronic Greens functions by several steps as in \disp{convolute}, and is therefore somewhat indirect. In contrast, 
 it has been pointed out recently  \cite{ARPES} that a  direct answer  of the Fermi-liquid- non Fermi-liquid question can be  found,  by using  the angle resolved photo emission  to experimentally infer the imaginary self-energy of the electron  from the spectral intensity near the Fermi wave vector. We hope that these results will provide motivation for carrying out further experiments suggested here.

}

\section{Supplemental Material}

For  convenience we created smaller $N_k=92$, $N_{\omega}=2^{12}$ files from which we fit and stored the spectral data as polynomials. We include this data as supplemental material, along with a Jupyter notebook for processing. This can be used to retrieve a reasonable approximation of our resistivities, to interpolate to new resistivities at different $n$ values and to perform any other desired calculations with the spectral functions. \textcolor{black}{To make the set of available data comprehensive we have also included BSLCO and LSCO results. See the README file for more information. The files can be found in Zenodo \href{https://doi.org/10.5281/zenodo.15306960}{here} \cite{data}.}

\section{Acknowledgments}

This work used Expanse by Dell and SDSC through allocation DMR170044 from the Advanced Cyberinfrastructure Coordination Ecosystem: Services and Support (ACCESS) program, which is supported by National Science Foundation grants 2138259, 2138286, 2138307, 2137603, and 2138296 \cite{XSEDE}. \sri{We thank Professor A. Carrington, Professor J. R. Cooper, Professor M. Greven, and an anonymous referee for helpful comments.}


\begin{thebibliography}{99}
\bibitem{expl-x}  For data from the Bi2201 family, the method adopted here for converting  $T_c$ to  x following \cite{Hussey} is as follows. The samples  are located on the $T_c$-$x$ parabolic plot assumed to be of the form \disp{Presland} \cite{Pres}.  This phenomenological relation   relates the hole doping $x$ to the transition temperature $T_c$ through 
\beq
x=0.16 \pm 0.11 \left\{1- \frac{T_c}{T_c^{Max}}\right\}^\half .\label{Presland}
\eeq  
Here  the system dependent $T_c^{Max}$ is taken to  be $\sim 35$K, and yields a superconducting dome in the (hole) density range $0.05 \leq x \leq 0.27$, with an optimal value $x_{opt}=0.16$.
\end{thebibliography}
\end{document}